\documentclass[preprint,prc,aps,epsfig]{revtex4}
\usepackage{psfig}
\def\beq{\begin{equation}}
\def\eeq{\end{equation}}
\def\beqa{\begin{eqnarray}}
\def\eeqa{\end{eqnarray}}

\def\GeV{\nobreak\,\mbox{GeV}}

\def\Tr{\mbox{ Tr }}

\def\me#1{\langle{#1}\rangle}

\def\bra#1{\langle #1|}
\def\ket#1{| #1\rangle}
\def\qbar{\overline{q}}
\def\gs{g_{\rm s}}
\def\G{{\cal G}}
\def\mixbar{\gs\qbar\sigma\!\cdot\!\G q}

\def\gluoncon{{\displaystyle{\gs^2G^2}}}
\def\qslash{\rlap{/}{q}}
\def\xsla{\rlap{/}{x}}
\def\qsq{q^2}

\begin{document}

\title{\sc A Comparative Study of Pentaquark Interpolating Currents}
\author{R.D. Matheus, F.S. Navarra, M. Nielsen, R. Rodrigues da Silva}
\affiliation{Instituto de F\'{\i}sica, Universidade de S\~{a}o Paulo\\
 C.P. 66318,  05315-970 S\~{a}o Paulo, SP, Brazil}

\begin{abstract}

In a diquark-diquark-antiquark picture of pentaquarks, we use two interpolating
currents to calculate the mass of the recently measured $\Xi^{--}$ state in the 
framework of QCD sum rules. We show that, even though yielding similar values 
for $m_{\Xi^{--}}$ (and close to the experimental value), these currents differ 
from each other in what concerns the strength of the pole, convergence of the 
OPE and sensitivity to the continuum threshold parameter.
\end{abstract}

\maketitle

\vspace{1cm}
\section{Introduction}

After the recent discovery of  pentaquark states  
\cite{lep,diana,clas,saphir,alt}  the central 
 question to be addressed now concerns the structure of these
new baryons. 
They could be: a) uncorrelated quarks inside a bag \cite{strot}; 
b) a $K-N$ molecule bound by a van-der Waals force \cite{brodsky}; c) a 
``$K-N$'' bound state in which $u u d$ and $u \overline{s}$ are not 
separately in color singlet states \cite{zhu}; d) a diquark-triquark 
$(u d) - (u d \overline{s})$ bound state \cite{kali} and 
e) a diquark-diquark-antiquark (DDA) state. This last one was 
advanced by  Jaffe and Wilczek (JW) \cite{jawil} and is quite appealing 
because it can explain  two unusual features of pentaquarks: their small 
mass and decay width.  Instantons generate strong attractive quark-quark 
interactions with  the formation of low mass diquarks, which, in turn,  lead to 
relatively low mass pentaquarks. This was verified in  \cite{mnnr}, \cite{oka} 
and \cite{eiden}. Moreover, in \cite{stech}, it was shown that 
the DDA configuration may  lead to strongly suppressed transition amplitudes (to 
meson-baryon states) for a reasonable choice of its spatial wave function, namely, 
two separated  extended diquark balls overlapping only partially and the antiquark 
at the center of the system. 

Pentaquark  configurations can be implemented in  QCD sum rules (QCDSR) \cite{svz,rry} 
and in lattice QCD \cite{latt,sasaki} by a proper choice of the interpolating current. 
A current for 
configuration c) has been proposed by Zhu \cite{zhu} and for configuration e) 
three different currents were proposed in \cite{mnnr}, \cite{oka} and 
\cite{eiden}. 

Given the impulse of this field and the prospects of new measurements, we may 
expect that the  ``pentaquark wave'' will last still for a long time. A great  
effort will be devoted to understand the structure of these objects. For the 
QCDSR community (and for lattice QCD studies as well) this means  that more 
attention will have to  be given to the properties of the interpolating 
currents.  At this point, we can already compare the three calculations presented 
in  \cite{mnnr}, \cite{oka} and \cite{eiden} for the mass of the $\Theta^{+}$. 
Although they present different implementations of the DDA scheme, they produce 
very similar  results for the pentaquark mass. This indicates that a 
more careful analysis has to be performed.

The purpose of the present work is twofold. We will use QCDSR to calculate the 
mass of the recently measured  $\Xi^{--}$ and we will also take the opportunity to 
perform a careful comparison of the results obtained with two different currents, 
which we will call I and II. Current I was introduced by  us in \cite{mnnr} and 
current II is the one proposed in \cite{oka}.  As it will be seen, both 
currents  give approximately the same mass for the $\Xi^{--}$ but have 
different sensitivity to the continuum threshold parameter, different convergence 
of the operator product expansion (OPE) and also different strength of the pole, 
with respect to the continuum. This comparative study, which will later be 
extended to the currents suggested in \cite{eiden} and also in \cite{gloz} , 
will give us a better understanding of these currents and  may  lead  to the 
choice of the best current.

\section{Currents and correlation functions}

In the QCDSR approach \cite{svz,rry}, the short range perturbative QCD is
extended by an OPE expansion of the correlators, which results in 
a series in powers of
the squared momentum with Wilson coefficients. The convergence at low
momentum is improved by using a Borel transform. The expansion involves
universal quark and gluon condensates. The quark-based calculation of
a given correlator is equated to the same correlator, calculated using
hadronic degrees of freedom via a dispersion relation, providing sum rules
from which a hadronic quantity can be estimated. The QCDSR 
calculation of hadronic masses centers around the two-point
correlation function given by
\beq
\Pi(q)\equiv i\int d^4 x\, e^{iq\cdot x}
\bra{0} T\eta(x)\overline{\eta}(0)\ket{0}\ ,
\label{cor}
\eeq
where $\eta(x)$ is an interpolating field (a current) with the quantum
numbers of the hadron we want to study.

\subsection{Current I}

Following the diquark-diquark-antiquark scheme, we can write two 
independent interpolating fields with the quantum numbers of $\Xi^{--}$: 
\beq
\eta_1(x)={1\over\sqrt{2}}\epsilon_{abc} (d_a^T(x) C \gamma_5 s_b(x))
[d_c^T(x) C \gamma_5 s_e(x) +   s_e^T(x) C \gamma_5 d_c(x)] C\bar{u}^T_e(x) 
\label{eta1}
\eeq
\beq
\eta_2(x)={1\over\sqrt{2}}\epsilon_{abc} (d_a^T(x) C  s_b(x))
[d_c^T(x) C  s_e(x) +   s_e^T(x) C  d_c(x)] C\bar{u}^T_e(x) 
\label{eta2}
\eeq
where $a,~b,~c$ and $e$ are color index and $C = - C^T$ is the charge 
conjugation operator. 

As in the nucleon case, where one also has two independent currents with 
the nucleon quantum numbers \cite{io1,dosch}, the most general current for
$\Xi^{--}$ is a linear combination of the currents given above:
\beq
\eta(x)=\left[t\eta_1(x) + \eta_2(x)\right] ,
\label{cur}
\eeq
with $t$ being an arbitrary parameter. In the case of the nucleon, the 
interpolating field with $t=-1$ is known as Ioffe's current \cite{io1}.
With this choice for $t$, this current maximizes the overlap with the nucleon 
as compared with the excited states, and minimizes the contribution of higher
dimension condensates. In the present case it is not clear a priori which 
is the best choice for $t$. This will be studied in what follows.

\subsection{Current II}

We also employ the following interpolating field operator for the pentaquark 
$\Xi^{--}$ \cite{oka,sasaki} :
\begin{eqnarray}
  \eta(x)&=&\epsilon^{abc}\epsilon^{def}\epsilon^{cfg}
   \{s_a^T(x)Cd_b(x)\}\{s_d^T(x)C\gamma_5 d_e(x)\}C\bar{u}_g^T(x)  
   \label{eq:IF}
\end{eqnarray}
It is easy to confirm that this operator produces a baryon with $J=1/2$, 
$I=3/2$ and strangeness $-2$.
The parts,
$S^c(x) = \epsilon^{abc} s_a^T(x)C\gamma_5 d_b(x)$ and
$P^c(x) = \epsilon^{abc} s_a^T(x)Cd_b(x)$,
give the scalar $S$ ($0^+$) and the pseudoscalar $P$ ($0^-$)  $ s d $ diquarks, 
respectively.  They both belong to the anti-triplet  representation of the color 
SU(3). 
The scalar diquark corresponds to the the $I=1/2$ $s d$ quark with zero angular 
momentum.  It is known that a gluon exchange force as well as the instanton 
mediated force commonly used in the quark model spectroscopy give significant 
attraction between the quarks in this channel.

\subsection{Mass sum rules for the $\Xi^{--}$}

Inserting Eq.~(\ref{cur}) into the integrand of Eq.~(\ref{cor}) we obtain
\beq
\bra{0} T\eta(x)\overline{\eta}(0)\ket{0}=
t^2\Pi_{11}(x) + t\left(\Pi_{12}(x)+\Pi_{21}(x)\right)+\Pi_{22}(x)\,
\eeq
Calling $\Gamma_1=\gamma_5$ and $\Gamma_2=1$ we get
\beqa
\Pi_{ij}(x)=\bra{0} T\eta_i(x)\overline{\eta}_j(0)\ket{0}=
\epsilon_{abc}\epsilon_{a'b'c'}C{S_{e'e}}^T(-x)C\left\{
-\Tr[\Gamma_iS^s_{bb'}(x)\Gamma_jCS^T_{aa'}(x)C]
\right.
\nonumber\\
\times\Tr[\Gamma_iS^s_{ee'}(x)\Gamma_jCS^T_{cc'}(x)C]
+\Tr[\Gamma_iS^s_{be'}(x)\Gamma_jCS^T_{cc'}(x)C\Gamma_iS^s_{eb'}(x)\Gamma_j
CS^T_{aa'}(x)C]
\nonumber\\
+\Tr[\Gamma_iS^s_{bb'}(x)\Gamma_jCS^T_{ca'}(x)C\Gamma_iS^s_{ee'}(x)\Gamma_j
CS^T_{ac'}(x)C]-\Tr[\Gamma_iS^s_{be'}(x)\Gamma_jCS^T_{ac'}(x)C]
\nonumber\\
\times\Tr[\Gamma_iS^s_{eb'}(x)\Gamma_jCS^T_{ca'}(x)C]
-\Tr[\Gamma_iS_{ab'}(x)\Gamma_jCS^{sT}_{ea'}(x)C\Gamma_iS_{ce'}(x)\Gamma_j
CS^{sT}_{bc'}(x)C]
\nonumber\\
-\Tr[\Gamma_iS^s_{ba'}(x)\Gamma_jCS^T_{cb'}(x)C\Gamma_iS^s_{ec'}(x)\Gamma_j
CS^T_{ae'}(x)C]+\Tr[\Gamma_iS^s_{ba'}(x)\Gamma_jCS^T_{ab'}(x)C]
\nonumber\\
\left.
\times\Tr[\Gamma_iS_{ce'}(x)\Gamma_jCS^{sT}_{ec'}(x)C]
+\Tr[\Gamma_iS^s_{bc'}(x)\Gamma_jCS^T_{ae'}(x)C]\Tr[\Gamma_iS^s_{ea'}(x)\Gamma_j
CS^T_{cb'}(x)C]\right\}\;,
\label{expa}
\eeqa
where $S_{ab}(x)$ and $S_{ab}^s(x)$ are the light and strange quark
propagators respectively.

In order to evaluate the correlation function $\Pi(q)$ at the quark
level, we first need to determine the quark propagator in the
presence of quark and gluon condensates. Keeping track of the terms
linear in the quark mass and taking into account quark and gluon 
condensates, we get \cite{yhhk}
\beqa
S_{ab}(x)&=&\bra{0} T[q_a(x)\overline{q}_b(0)]\ket{0}={i\delta_{ab}\over2
\pi^2x^4}\xsla-{m_q\delta_{ab}\over4\pi^2x^2}-{i\over32\pi^2x^2}t^A_{ab}
\gs G^A_{\mu\nu}
(\xsla\sigma^{\mu\nu}+\sigma^{\mu\nu}\xsla)
\nonumber\\
&-&{\delta_{ab}\over12}\me{\qbar q}
-{m_q\over32\pi^2}t^A_{ab}\gs G^A_{\mu\nu}\sigma^{\mu\nu}\ln(-x^2)
+{i\delta_{ab}\over48}m_q\me{\qbar q}\xsla-{x^2\delta_{ab}
\over2^6\times3}\me{\mixbar}
\nonumber\\
&+&{ix^2\delta_{ab}\over2^7\times3^2}m_q\me{\mixbar}\xsla
-{x^4\delta_{ab}\over2^{10}\times3^3}\me{\qbar q}\me{\gluoncon}\,,
\label{prop}
\eeqa
where we have used the factorization approximation for the multi-quark
condensates, and we have used the fixed-point gauge \cite{yhhk}.

Inserting (\ref{prop}) into (\ref{expa}) we obtain a set of diagrams which 
contribute to the OPE side of the correlation function. In the case of the 
current I, Fig. 1 shows the classes of diagrams which we are considering. 
In  each diagram, except the perturbative one, 
all possible permutations of the lines are implicit.

Lorentz covariance, parity and time reversal imply that the two-point
correlation function in Eq.~(\ref{cor}) has the form
\beq
\Pi(q)= \Pi_1(q^2)+ \Pi_q(\qsq) \qslash\;.
\label{stru}
\eeq
A sum rule for each scalar invariant function $\Pi_1$ and $\Pi_q$, can be 
obtained. As in ref.~\cite{zhu}, in this work we shall use the chirality
even structure $\Pi_q(\qsq)$ to obtain the final results but we shall also 
discuss the chirality odd structure $\Pi_1$, showing its limitations for 
our present purposes.

The phenomenological side is described, as usual, as a 
sum of pole and continuum, the latter being approximated  by the OPE 
spectral density.
In order to suppress the condensates of higher dimension and at the same time
reduce the influence of higher resonances  we perform a  
standard Borel transform \cite{svz}: 
\beq
\Pi (M^2) \equiv \lim_{n,Q^2 \rightarrow \infty} \frac{1}{n!} (Q^2)^{n+1} 
\left( - \frac{d}{d Q^2} \right)^n \Pi (Q^2)
\label{borel}
\eeq
($Q^2 = - q^2$) with the squared Borel mass scale $M^2 = Q^2/n$ kept 
fixed in the limit.

For current II we repeat the steps mentioned above, substituting (\ref{eq:IF})
into (\ref{cor}), making use of the expansion (\ref{prop}), picking up the terms 
multiplying the structure $ \qslash  $ and finally performing a Borel transform 
(\ref{borel}). In this case the diagrams that have to be considered are shown in
Fig. 2.

After Borel transforming each side of $\Pi_q(Q^2)$ and transferring
the continuum contribution to the OPE side we obtain the following
sum rule at order $m_s$:
\begin{equation}
\label{soma}
\lambda_{\Xi}^{2} \, e^{-\frac{m_{\Xi}^{2}}{M^{2}}}
=
\int_{0}^{s_0}e^{-\frac{s}{M^2}}
\rho^q_{i}(s)ds.
\end{equation}
where $i (=I,II)$ refers to the current employed and the spectral densities, up 
to order 6 are given by:
\begin{eqnarray}
\label{rho1}
\rho^q_I \, (s) &=&c_1\frac{s^5}{5!5!2^{12}7\pi^{8}}+
c_3\frac{s^3}{5!2^{10}\pi^{6}}m_s<\bar{s}s>-
c_4\frac{s^3}{5!2^{8}\pi^{6}}m_s<\bar{q}q>\nonumber\\
&+&
c_2\frac{s^3}{5!3!2^{13}\pi^{6}}<\frac{\alpha_s}{\pi}G^2>+
7c_4\frac{s^2}{2^{16}\pi^{6}}m_s<\bar{q}g_s{\bf \sigma.G}q>\nonumber\\
&+&
c_2\frac{s^2}{3^{2}2^{11}\pi^{4}}\left(<\bar{s}s>^2+<\bar{q}q>^2\right)+
c_4\frac{s^2}{3!2^{8}\pi^{4}}<\bar{s}s><\bar{q}q>\nonumber\\
&-&c_5\frac{s^2}{4!3!2^{11}\pi^{6}}m_s<\bar{s}g_s{\bf \sigma.G}s>\nonumber\\
&-&
3 c_4\frac{s^2}{4!3!2^{12}\pi^{6}}m_s<\bar{q}g_s{\bf \sigma.G}q>
\left(6\, \mbox {ln}(\frac{s}{\Lambda^2_{QCD}})-\frac{43}{2}\right),\nonumber\\
\end{eqnarray}
with
$c_1=5t^2+2t+5$, $c_2=(1-t)^2$, $c_3=(t+1)^2$, $c_4=t^2-1$, 
$c_5=t^2+22t+1$, $\Lambda_{QCD} = 110$ MeV 
and 
\begin{eqnarray}
\label{rho2}
\rho^q_{II} \, (s) &=&\frac{s^5}{5!5!2^{10}7\pi^{8}}+
\frac{s^3}{5!3!2^{7}\pi^{6}}m_s<\bar{s}s>+
\frac{s^3}{5!3!2^{10}\pi^{6}}<\frac{\alpha_s}{\pi}G^2>\nonumber\\
&+&\frac{s^2}{4!3!2^{9}\pi^{6}}m_s<\bar{s}g_s{\bf\sigma.G}s>.
\end{eqnarray}
To extract the $\Xi^{--}$ mass, $m_{\Xi}$, we take the derivative of 
Eq.~(\ref{soma}) with respect to $M^{-2}$ and divide it by Eq.~(\ref{soma}). 
Repeating the same steps leading to (\ref{soma}),  (\ref{rho1}) and (\ref{rho2}) 
for the chirality odd structure  $\Pi_1(q^2)$ we arrive at 
\begin{equation}
\label{soma}
\lambda_{\Xi}^{2} \, m_{\Xi} \, e^{-\frac{m_{\Xi}^{2}}{M^{2}}}
=
\int_{0}^{s_0}e^{-\frac{s}{M^2}}
\rho^1_{i}(s)ds.
\end{equation}
where 
\begin{equation}
\rho^1_I \, (s) \, = \, -c_1\frac{s^4}{5!4!2^{10}\pi^{6}}<\bar{q}q> \, \, + \, \,
c_1\frac{s^3}{4!3!2^{12}\pi^{6}}<\bar{q}g_s{\bf \sigma.G}q> 
\label{rho11}
\end{equation}
and
\begin{equation}
\rho^1_{II} \, (s) \, = \, -\frac{s^4}{5!4!2^{7}\pi^{6}}<\bar{q}q> \, \, + \, \,
\frac{s^3}{4!3!2^{9}\pi^{6}}<\bar{q}g_s{\bf \sigma.G}q> 
\label{rho21}
\end{equation}

In the complete theory,  the mass extracted from the sum rule
should be independent of the Borel mass $M^2$. However, in  a truncated 
treatment there will 
always  be some dependence left.  Therefore, one has to work in a region 
where the approximations made are supposedly acceptable and where 
the result depends only moderately on the Borel variables.

\section{Evaluation of the sum rules and results}

Before proceeding with the numerical analysis, a remark is in order. A 
comparison between results obtained with different currents is more  meaningful 
when they describe  the same physical state, i.e., those with the same quantum 
numbers. Concerning spin, all currents considered in our work have the same spin 
(= 1/2). Concerning the parity, the situation is more complicated. In QCD sum 
rules, when we construct the current, it has a definite parity.  
Current I has parity $P=-1$ and current II has parity $P=+1$. However,  currents 
can couple to physical states of different parities.  As well discussed in 
\cite{suhong_c}, in order to know the parity of the state in QCDSR, we have to 
analyze the chiral-odd sum rule. If the r.h.s of this sum rule (containing the 
spectral density coming from QCD) is positive, then the parity of the corresponding 
physical state is the same as the parity of the current. If it is negative the parity 
is the opposite of the parity of the current.  Performing this analysis we might 
determine in both cases the parity of the state. However it turns out that for both 
currents, in the chiral-odd sum rule  the OPE does not have good convergence, i.e.,  
terms containing higher order operators are not suppressed with respect to the lowest 
order ones. This sum rule is thus ill defined and nothing can be said about the 
parity of the state. The comparative study of the currents is still valid because 
we can compare other properties of these currents. 
A possible outcome of this study might be that one current has defects, 
which are so severe that we are forced to abandon it. If this turns out to be case, 
the determination of the parity of the associated states becomes irrelevant.

Having presented the main formulas in the last section, the next step will be to 
introduce numerical values for the masses and condensates, choose reasonable values 
for the free (or partially constrained) parameters, which are the continuum 
threshold, the  Borel mass at which the  mass sum rule is evaluated and, 
in the case of current I, 
the value of $t$. As a result we obtain values for $m_{\Xi}$. In doing these 
calculations we must remember that {\it it is not enough to obtain a pentaquark 
mass consistent with the experimental number}. There is a list of requirements 
that must be fulfilled: \\
i) the physical observables, such as masses and coupling constants, must be
approximately independent of the Borel mass (this is the so called Borel 
stability). \\
ii) the right hand side (RHS) of the sum rule (\ref{soma}) must be positive, 
since the left hand side (LHS) is manifestly positive.\\
iii) the operator product expansion (OPE) must be convergent, i.e., the terms 
appearing in (\ref{rho1}) and (\ref{rho2}) must decrease with the increasing 
order of the operator. \\
iv) the pole contribution must be dominant, i.e., the integral in  (\ref{soma}) 
must be at least 50 \% of the  integral over the complete domain of invariant 
masses ($ s_0 \rightarrow \infty$). \\
v) the threshold parameter $s_0$ must be compatible with the energy corresponding 
to the first excitation of an usual baryon. \\
vi) the  current-state overlap $\lambda_{\Xi}$ must be sizeable. The larger is 
$\lambda_{\Xi}$ th better is the current.

As it will be seen the above conditions are not easy to be simultanously 
satisfied and may be useful in discriminating between different currents.

In the numerical analysis of the sum rules, the values used for the 
condensates are:  $\me{\qbar q}=\,
-(0.23)^3\,\GeV^3$,
$\langle\overline{s}s\rangle\,=0.8 \me{\qbar q}$, $ <\bar{s} g_s {\bf\sigma.G}s > 
= m_0^2 \me{\bar{s}s}$ with $m_0^2=0.8\,\GeV^2$ and $\me{\gluoncon}=0.5~\GeV^4$.
We define the continuum threshold as:
\beq
s_0 = (1.86+\Delta)^2   \,\,\,\, \mbox{GeV}^2
\label{s0}
\eeq

\subsection{current I}

We evaluate our sum rules in 
the range $1.0 \leq M^2 \leq 4.0 \GeV^2$. The results are shown in Table I. 
The first column in Table I gives the values of $t$ considered here. We notice
that some values were excluded, as for example $t=0$, because they would lead to 
a violation of condition ii) above. The simbol S stands for a current composed by
scalar diquarks only, i.e., $\eta_1$. The motivation for studying this current is
to verify if it gives a smaller mass for the pentaquark than those obtained with 
other currents. According to the instanton description of diquark dynamics, this 
should be the case. In the second and third columns of Table I we list the values 
adopted for the strange quark mass and continuum threshold respectively. The fourth 
column shows $m_{\Xi}$ and the Borel mass squared at which the particle mass was 
extracted. The fifth and sixth columns show the ratios of terms containing 
dimension 4 (quark and gluon condensates) and 6 (mixed condensates) 
operators and the perturbative term respectively. Ideally these 
numbers  should be smaller than one, the second being smaller than the first. Finally, 
the last column shows the strength of the pole.  In practice this quantity is obtained 
dividing the  
integral in the right hand side of (\ref{soma}) by the same integral with the 
upper limit $s_0 = \infty$. This percentage must be as large as 
possible, but usually we accept values around 50 \%.

As it can be seen in  Table I, it is hard to satisfy conditions i) - vi) 
simultaneously. In particular, when we have a very good OPE convergence, the 
strength of the pole is very weak and vice versa. We have to look for a 
compromise and we believe that the choices indicated in bold face are the best. 
For same choices for $m_s$ and $\Delta$ we observe that the masses found with 
scalar diquark currents, S, are only slightly smaller than  the others. This means
either that instanton dynamics was not captured by our choice of currents and 
diagrams or that the interaction between the pseudoscalar diquarks (included in
the mixed currents) is more attractive than expected. The connection between our
approach and instantons deserves further investigation. 

In order to further illustrate these results, we consider the parameters of the 
first line in bold face, $m_s=0.10$ GeV and $\Delta=0.44$ GeV, 
and contruct, Figures 3, 4 and 5, showing the Borel mass 
dependence of  $m_{\Xi}$, of the OPE terms (in absolute value) and  of the  
percentage of the pole contribution  respectively. For these choices the value 
of the current-state overlap is:
\beq
\lambda^q _{I} \, \simeq \, 5.4 \,  \times \, 10^{-9}  \,\,\,\, \mbox{GeV$^{13}$}
\label{lambnos}
\eeq
In view of the variety of results presented in Table I and in the figures, it 
would be somewhat artificial to invent a procedure to determine the average value
of the $\Xi$ mass and its corresponding error. If we simply take the average over 
all values presented we arrive at:
\beq
 m_{\Xi} = 1.87 \pm 0.22   \,\,\,\, \mbox{GeV}
\label{mIa}
\eeq
Restricting ourselves to the parameter combinations which satisfy the requirements
i) - vi) and taking the average we obtain:
\beq
 m_{\Xi} = 1.85 \pm 0.05   \,\,\,\, \mbox{GeV}
\label{mIb}
\eeq

These numbers are very close to the experimental value. 

\subsection{current II}

Using the same numerical inputs quoted in the last subsection we  evaluate now the 
sum rules obtained with current II. The results are shown in Table II, which has the 
same format as Table I except for the fact that,now,  there is no $t$ parameter. 
The same comments made for the results of Table I apply here. Choosing the 
parameters of the first line in bold face, $m_s=0.10$ GeV and $\Delta=0.24$ GeV, we 
present  in Figures 6, 7 and 8 the Borel mass dependence 
of  $m_{\Xi}$, of the OPE terms and  of the  percentage of the pole respectively.
As it can be seen, with current II we tend to overestimate $m_{\Xi}$, unless very 
low threshold parmeters or quark masses are used. Besides, the pole contribution is 
always smaller than 40 \%.  On the other hand, the Borel stability seen in Fig. 6 
is remarkable. This suggests that we could choose a lower value for the Borel mass, 
thereby increasing the pole contribution without significantly changing  $m_{\Xi}$. 
In order to extract from Table II some value for the pentaquark mass, we repeat the 
procedure applied in the last subsection. The simple average gives:
\beq
 m_{\Xi} = 2.00 \pm 0.16  \,\,\,\, \mbox{GeV}
\label{mIIa}
\eeq
and the average over the best parameter choices leads to:
\beq
 m_{\Xi} = 1.88 \pm 0.04   \,\,\,\, \mbox{GeV}
\label{mIIb}
\eeq
These numbers are somewhat high but still compatible with data. Finally, 
for these parameters the current-state overlap is:
\beq
\lambda^q _{I} \, \simeq \, 1.3 \,  \times \, 10^{-9}  \,\,\,\, \mbox{GeV$^{13}$}
\label{lamboka}
\eeq

Comparing (\ref{lambnos}) and (\ref{lamboka}) we observe that the coupling of 
current I to the the $\Xi$ state is four times larger than the coupling of 
current II to this state. This speaks in favor of current I.

\section{Summary and conclusions}

In conclusion, we have presented a QCD sum rule study of the $\Xi^{--}$
pentaquark mass using a  diquark-diquark-antiquark scheme. We have 
employed two possible currents (I and II) which implement this picture. Several 
parameter choices were considered  for the currents. We could conclude that for 
both currents it was possible to find a set of parameters and a Borel window so 
that conditions i) - vi) were fulfilled and a reasonable value for the mass was 
found. The ratio pole/total, presented in the tables is still small and probably 
is the most important drawback of the studied currents. 

At the present stage we may say that the pentaquark  $\Xi^{--}$ can be reasonably  
well described by QCD sum rules and the value of its mass is given by (\ref{mIb}) 
or by (\ref{mIIb}). These numbers can be considered the same within the uncertainties. 
Concerning the currents, there are  differences in the Borel stability, 
in the sensitivity to the strange quark mass, in the continuum threshold and in the
their couplings to the $\Xi$ state.

\underline{Acknowledgements}: 
We are indebted to S. Narison for fruitful discussions. 
 This work has been supported by CNPq and  FAPESP (Brazil). 


\eject

\begin{center}
\begin{tabular}{|c|c|c|c|c|c|c|}  \hline
t & $m_s$ $(GeV)$ & $\Delta$ (GeV) & $ m_{\Xi}(M_B^2)$ (GeV) & Dim 4/Pert
& Dim 6/Pert & Pole/Total \\
\hline
\hline
-1 & 0.15 & 0.44 & 1.95  (3.5) & 2.42 & 3.67  & 2.1  \% \\
\hline
-1 & 0.15 & 0.44 & 1.66  (1.0) & 0.36  & 7.7  & 80   \% \\
\hline
1 &  0.15 & 0.44 & 2.09  (3.5) & 0.96  & 0.53 & 0.28 \%  \\
\hline
1 &  0.15 & 0.44 & 1.82  (1.0) & 1.44  & 1.11 & 40   \%  \\
\hline
S & 0.15 & 0.44 & 1.94  (3.5) & 2.42  & 11.6 & 4.8  \% \\
\hline
S & 0.15 & 0.44 & 1.65  (1.0) & 3.62  & 23.9 & 87   \% \\
\hline
\hline
-1 & 0.12 & 0.44 & 1.95  (3.5) & 0.24  & 3.8  & 2.1  \% \\
\hline
-1 & 0.12 & 0.44 & 1.66  (1.0) & 0.36  & 8.0  & 80   \% \\
\hline
1 &  0.12 & 0.44 & 2.09  (3.5) & 0.76  & 0.42 & 0.32 \%  \\
\hline
1 &  0.12 & 0.44 & 1.86  (1.0) & 1.15  & 0.89 & 41   \%  \\
\hline
S & 0.12 & 0.44 & 1.94  (3.5) & 1.95  & 11.6 & 4.8  \% \\
\hline
S & 0.12 & 0.44 & 1.65  (1.0) & 2.93  & 24.1 & 87   \% \\
\hline
\hline
-1 & 0.10 & 0.44 & 1.95  (3.5) & 0.24  & 3.89  & 2.1  \% \\
\hline
-1 & 0.10 & 0.44 & 1.66  (1.0) & 0.36  & 8.2   & 80   \% \\
\hline
1 &  0.10 & 0.44 & 2.09  (3.5) & 0.64 & 0.35 & 0.34 \%  \\
\hline
{\bf 1} &  {\bf 0.10} & {\bf 0.44} & {\bf 1.88  (1.0)} &{\bf 0.96} 
&{\bf  0.74} & {\bf 41  \%}  \\
\hline
S & 0.10 & 0.44 & 1.93  (3.5) & 1.64 & 11.7 & 4.8  \% \\
\hline
S & 0.10 & 0.44 & 1.64  (1.0) & 2.47 & 24.3 & 87   \% \\
\hline
\hline
1 &  0.09 & 0.44 & 2.09  (3.5) & 0.57 & 0.31 & 0.36 \%  \\
\hline
{\bf 1} &  {\bf 0.09} & {\bf 0.44} & {\bf 1.90  (1.0)} &{\bf 0.86} 
&{\bf  0.67} & {\bf 42  \%}  \\
\hline
1 &  0.09 & 0.34 & 2.00  (3.5) & 0.68 & 0.40 & 0.23 \%  \\
\hline
1 &  0.09 & 0.34 & 1.81  (1.0) & 0.97 & 0.78 & 34   \%  \\
\hline
\hline
\end{tabular}
\end{center}
\begin{center}
\small{{\bf Table I:} Results for the $\Xi^{--}$ obtained 
with current I. } \\
\end{center}

\begin{center}
\begin{tabular}{|c|c|c|c|c|c|}  \hline
 $m_s$ (GeV) & $\Delta$ (GeV) & $ m_{\Xi}(M_B^2)$ (GeV) & Dim 4/Pert & Dim 6 /Pert
 & Pole/Total (\%) \\
\hline
\hline
0.15 & 0.24 & 2.03 $(3.5)$ & 0.02 & 0.44  & 0.11 \% \\
\hline
0.15 & 0.44 & 2.16 $(3.5)$ & 0.02 & 0.26  & 0.35 \%  \\
\hline
0.12 & 0.14 & 1.88 $(3.5)$ & 0.35 & 0.46  & 0.10 \% \\
\hline
0.12 & 0.14 & 1.93 $(1.0)^2 $ & 0.46 & 0.76  & 16 \% \\
\hline
0.10 & 0.24 & 1.93 $(3.5)$ & 0.47 & 0.29  & 0.22 \%  \\
\hline
{\bf 0.10} & {\bf 0.24} & {\bf 1.87  $(1.0)^2$} &{\bf 0.64} & {\bf 0.52}  
& {\bf  28.7 \% } \\ 
\hline
0.09 & 0.24 & 1.84  $(1.0)^2$ & 0.76 & 0.46  & 31.1 \%  \\
\hline
{\bf 0.09} & {\bf 0.34} &{\bf 1.91  $(1.0)^2$} &{\bf 0.67} & {\bf 0.39}  
& {\bf  39.0 \% } \\ 
\hline
\hline
\end{tabular}
\end{center}
\begin{center}
\small{{\bf Table II:} Results for the $\Xi^{--}$ obtained 
with current II. } \\
\end{center}

\begin{figure} \label{fig1}
\centerline{\psfig{figure=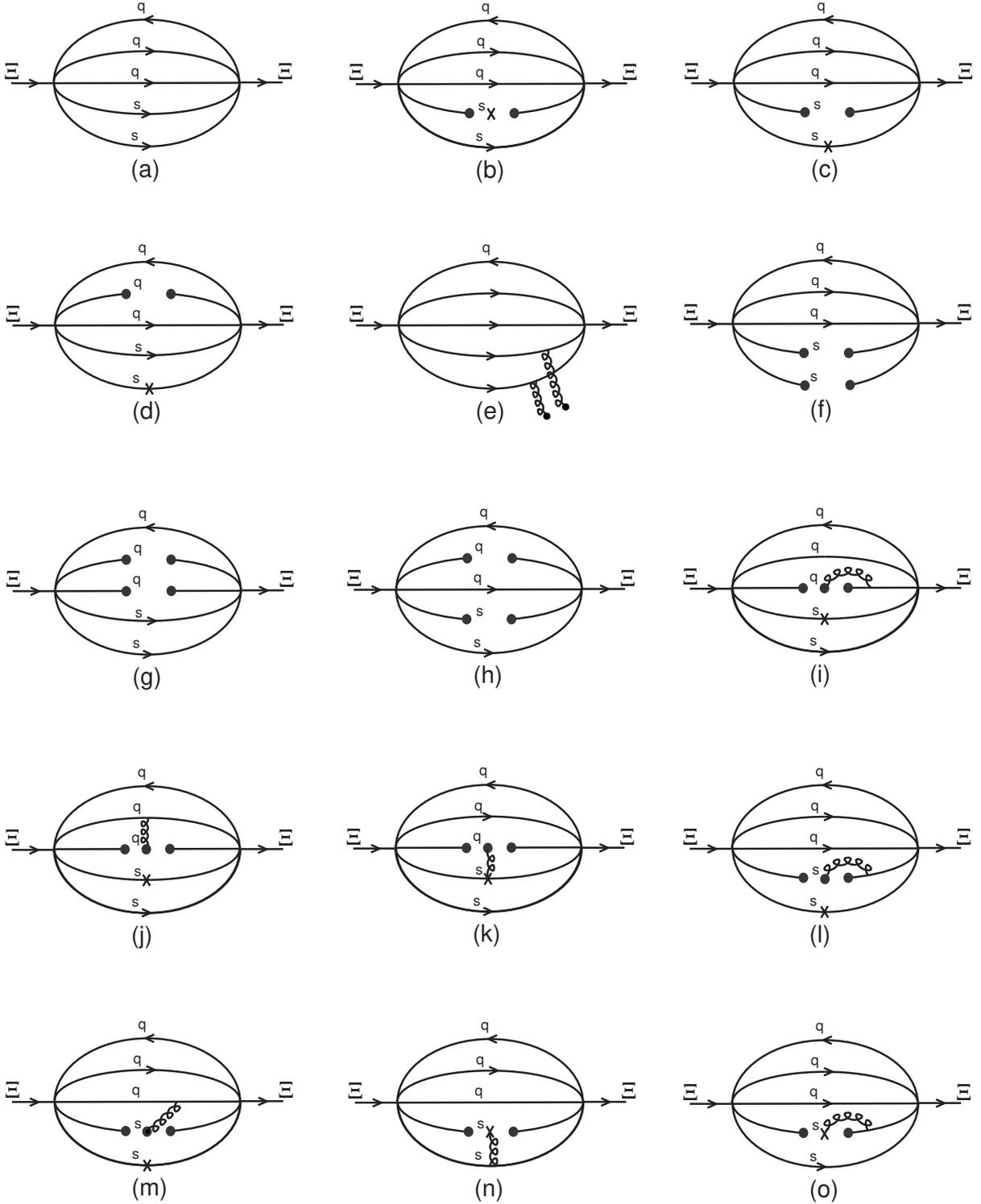,width=16cm,angle=0}}
\caption{Classes of diagrams which contribute to the OPE of current I up to 
dimension 6: a) perturbative; b)-d) quark condensate; e) gluon condensate; 
f)-h) 4-quark condensate; i)-o) mixed condensate.}
\end{figure}

\begin{figure} \label{fig2}
\centerline{\psfig{figure=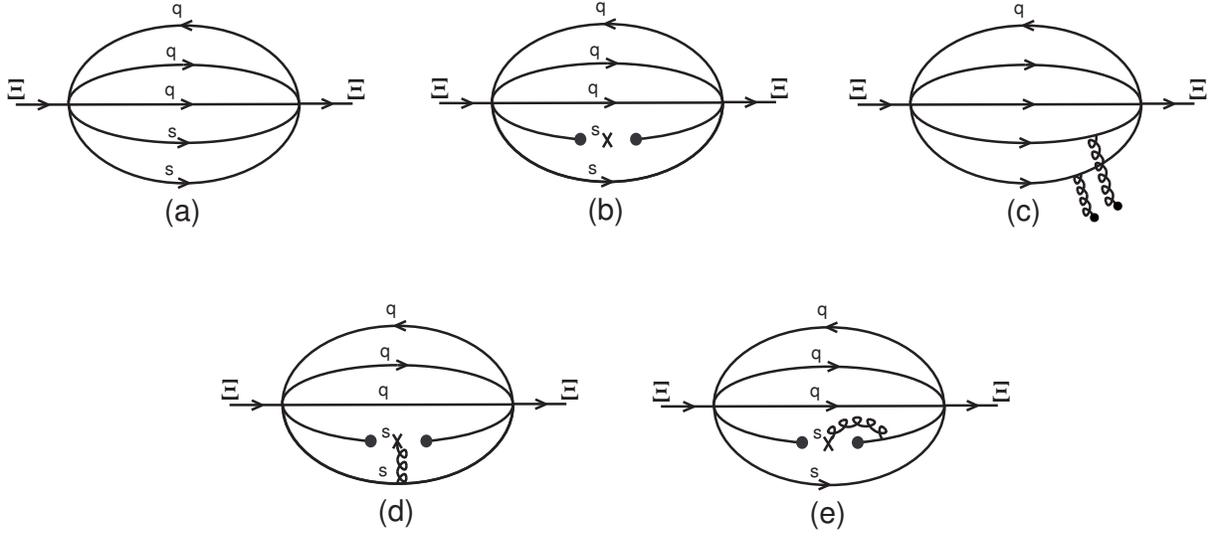,width=16cm,angle=0}}
\caption{Diagrams which contribute to the OPE of current II: a) perturbative;
b) quark condensate; c) gluon condensate; d)-e)  mixed condensate.}
\end{figure}

\begin{figure} \label{fig3}
\centerline{\psfig{figure=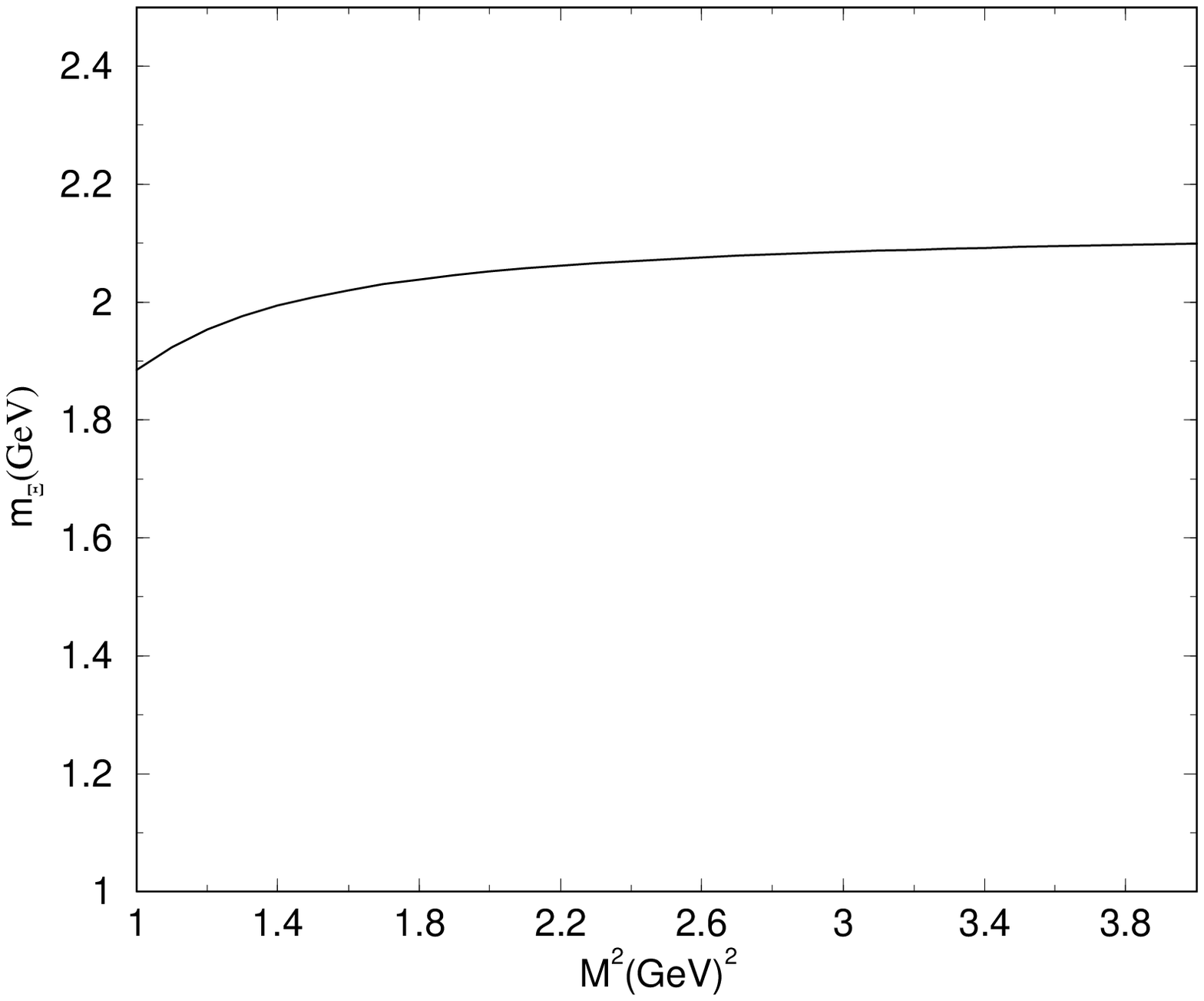,width=10cm,angle=0}}
\caption{$\Xi$ mass with current I. $m_s=0.10 \, \mbox{GeV}$, 
$t=1$  and $\Delta=0.44$ GeV.}
\end{figure}

\begin{figure} \label{fig4}
\centerline{\psfig{figure=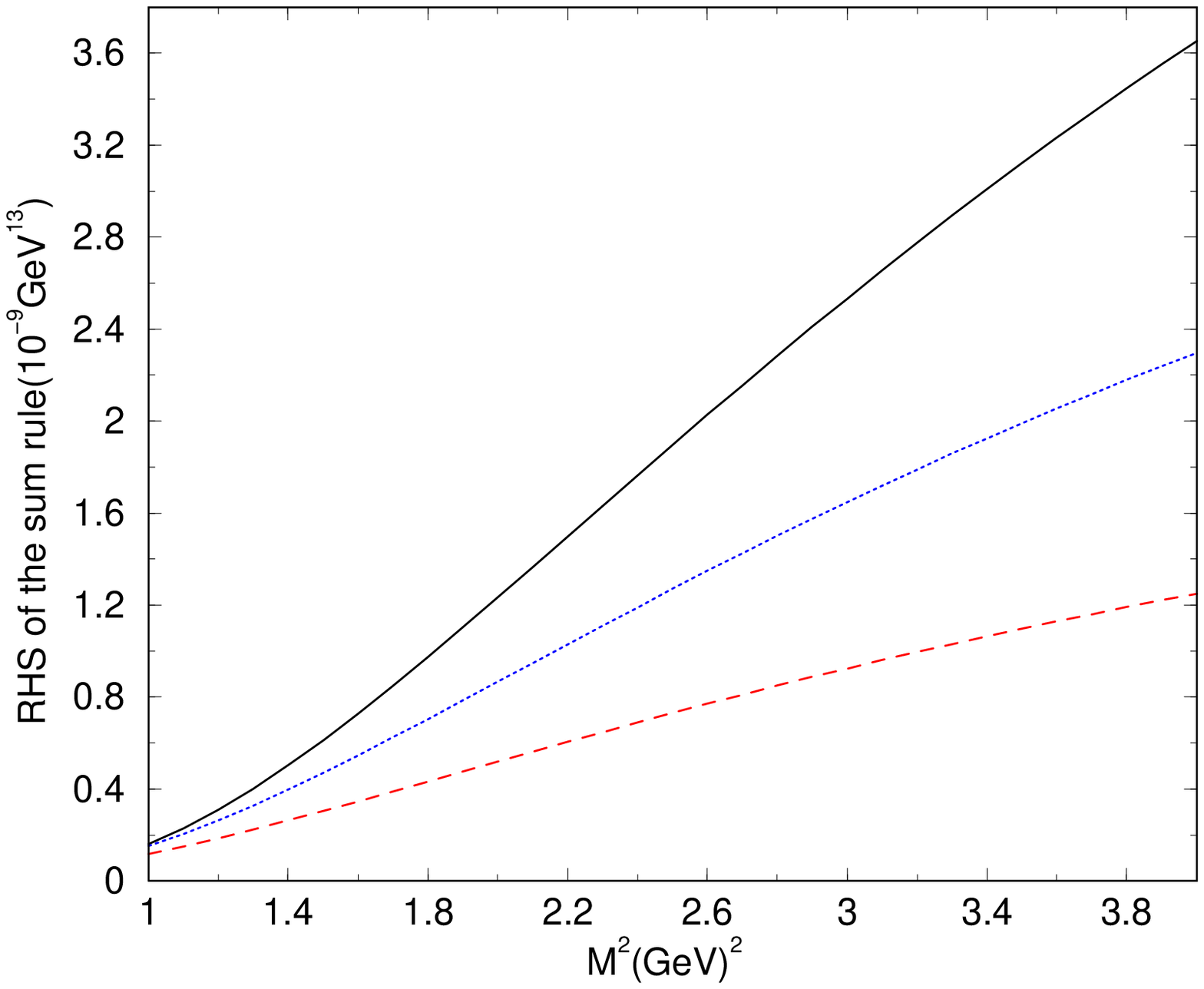,width=10cm,angle=0}}
\caption{Leading terms of the R.H.S of (\ref{soma}) with current I.
$m_s=0.10 \, \mbox{GeV}$, $t=1$  and $\Delta=0.44$ GeV.
Solid line: perturbative term; dotted line: operators of dimension 4; 
dashed line:operators of dimension 6.}
\end{figure}

\begin{figure} \label{fig5}
\centerline{\psfig{figure=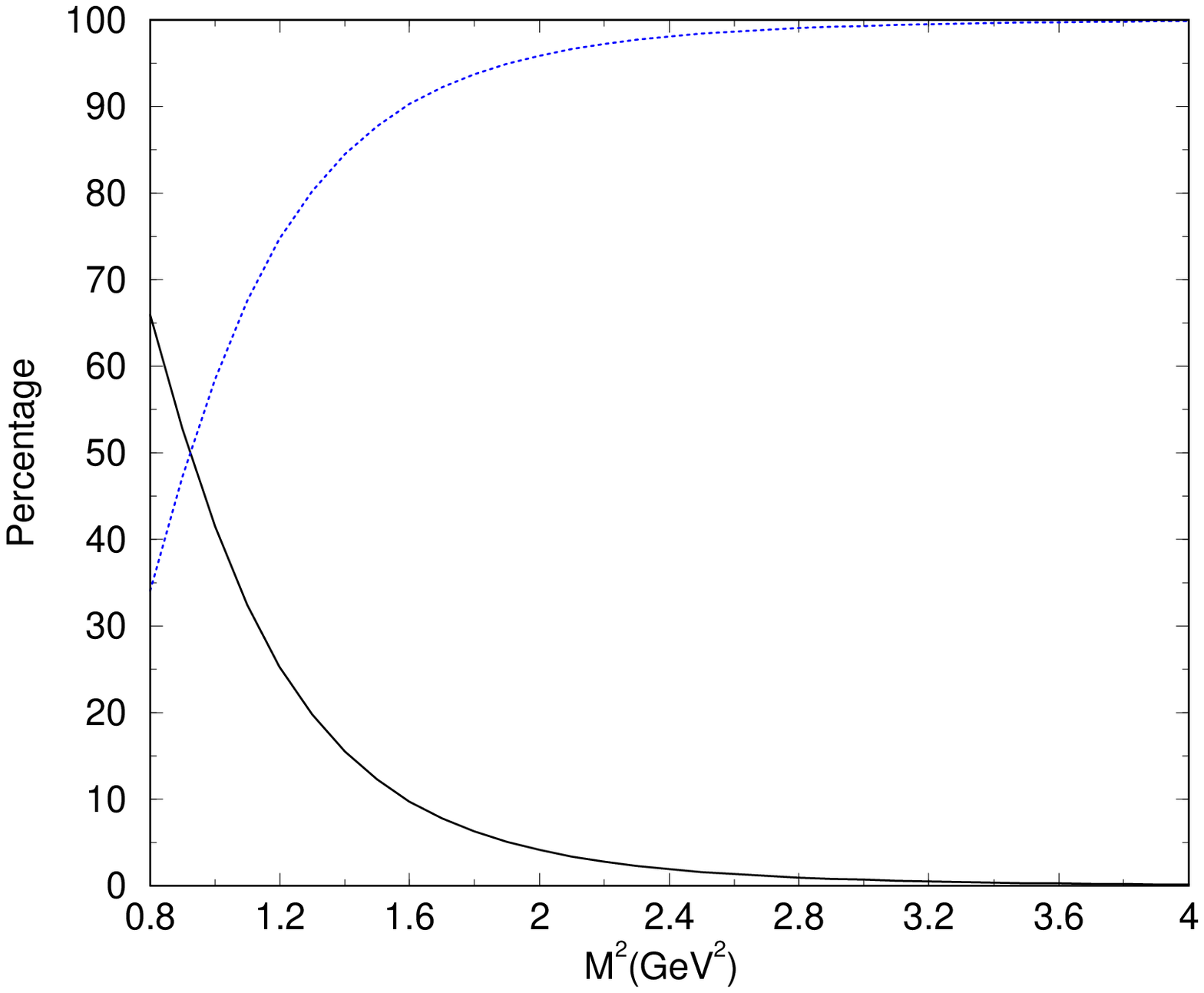 ,width=10cm,angle=0}}
\caption{Relative strength of the pole (solid line) and the continuum
(dotted line) as a function of the Borel mass squared with current I.
$m_s=0.10 \, \mbox{GeV}$, $t=1$  and $\Delta=0.44$ GeV.}

\end{figure}


\begin{figure} \label{fig6}
\centerline{\psfig{figure=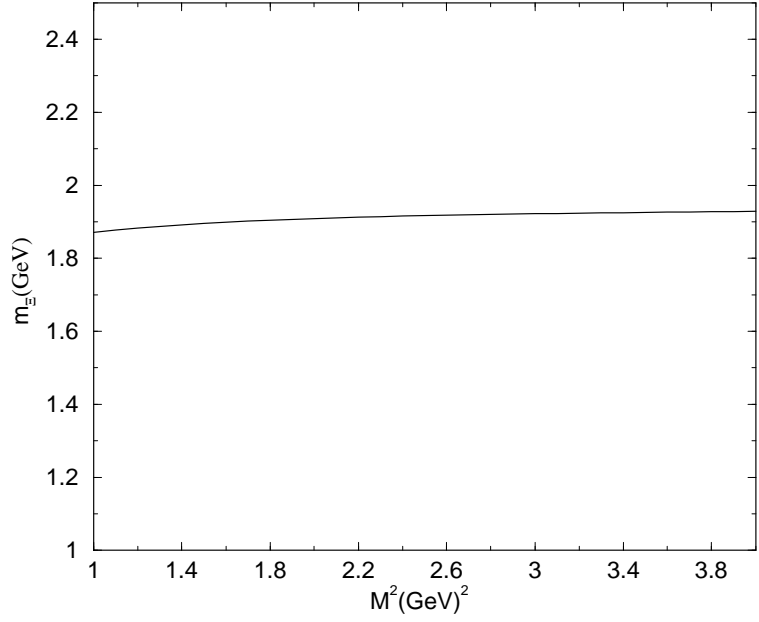,width=10cm,angle=0}}
\caption{$\Xi$ mass with current II. $m_s=0.10 \, \mbox{GeV}$ 
and $\Delta=0.24$ GeV.}
\end{figure}

\begin{figure} \label{fig7}
\centerline{\psfig{figure=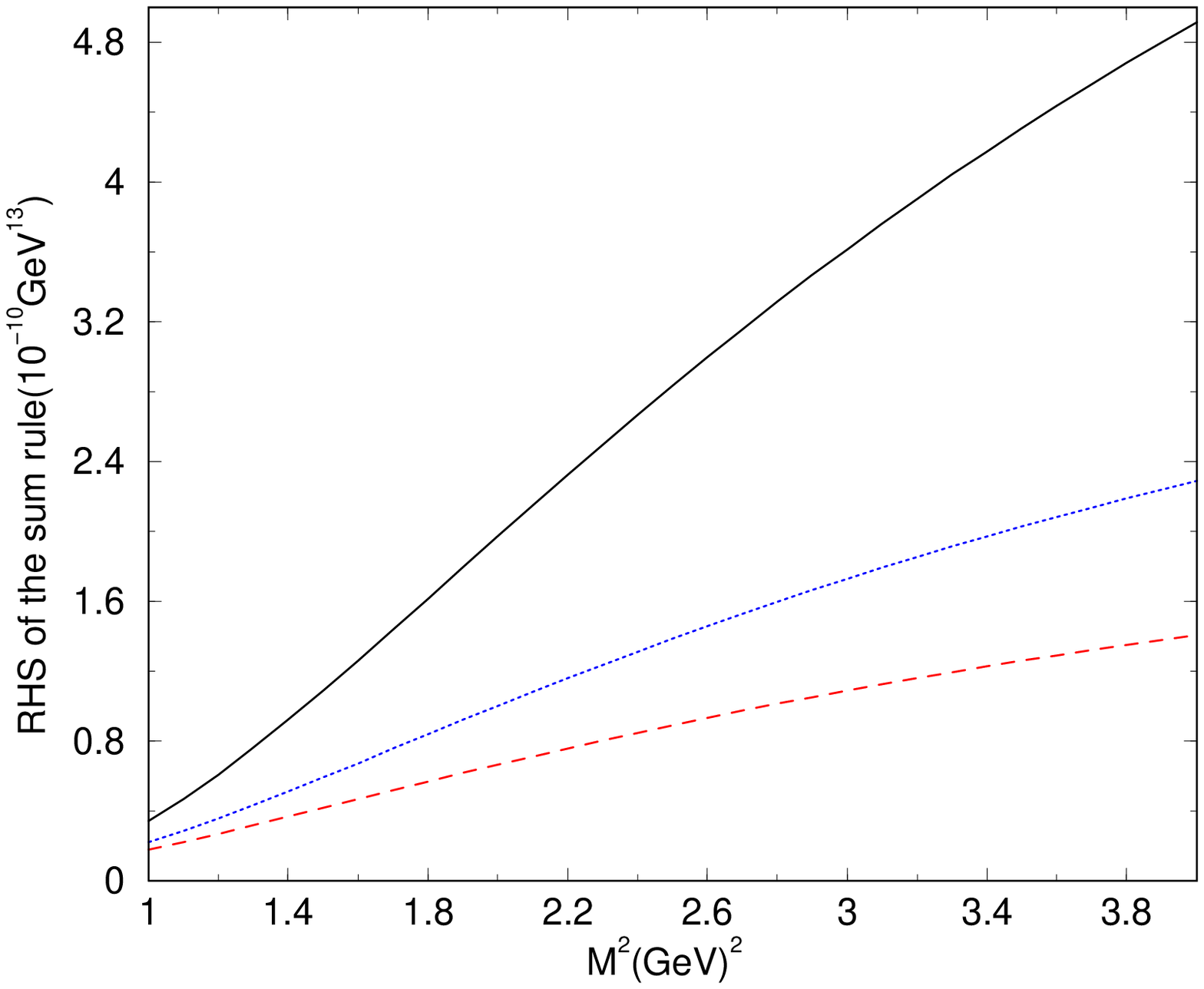,width=10cm,angle=0}}
\caption{Leading terms of the R.H.S of (\ref{soma}) with current II.
$m_s=0.10 \, \mbox{GeV}$   and $\Delta=0.24$ GeV.
Solid line: perturbative term; dotted line: operators of dimension 4; 
dashed line:operators of dimension 6.}

\end{figure}

\begin{figure} \label{fig8}
\centerline{\psfig{figure=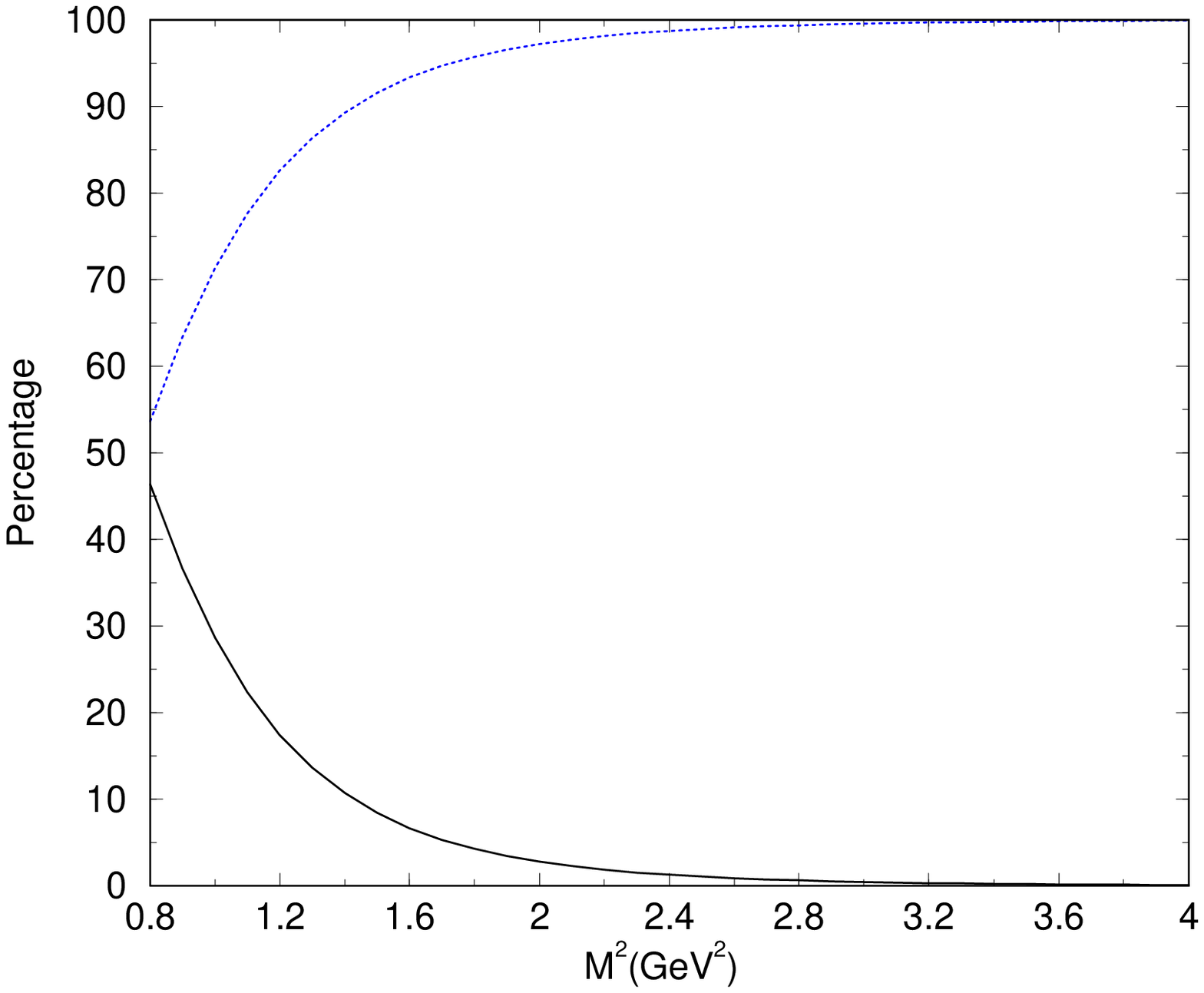 ,width=10cm,angle=0}}
\caption{Relative strength of the pole (solid line) and the continuum
(dotted line) as a function of the Borel mass squared with current II.
$m_s=0.10 \, \mbox{GeV}$, $t=1$  and $\Delta=0.24$ GeV.}
\end{figure}

\end{document}